\documentstyle[12pt]{article}

\catcode`\@=11
\long\def\@makefntext#1{
\protect\noindent \hbox to 3.2pt {\hskip-.9pt
$^{{\ninerm\@thefnmark}}$\hfil}#1\hfill}                

\def\@makefnmark{\hbox to 0pt{$^{\@thefnmark}$\hss}}  

\def\ps@myheadings{\let\@mkboth\@gobbletwo
\def\@oddhead{\hbox{}
\rightmark\hfil\ninerm\thepage}
\def\@oddfoot{}\def\@evenhead{\ninerm\thepage\hfil
\leftmark\hbox{}}\def\@evenfoot{}
\def\sectionmark##1{}\def\subsectionmark##1{}}

\renewcommand{\thefootnote}{\fnsymbol{footnote}}

\newcounter{sectionc}\newcounter{subsectionc}\newcounter{subsubsectionc}
\renewcommand{\section}[1] {\vspace*{0.6cm}\addtocounter{sectionc}{1}
\setcounter{subsectionc}{0}\setcounter{subsubsectionc}{0}\noindent
        {\normalsize\bf\thesectionc. #1}\par\vspace*{0.4cm}}
\renewcommand{\subsection}[1] {\vspace*{0.6cm}\addtocounter{subsectionc}{1}
        \setcounter{subsubsectionc}{0}\noindent
        {\normalsize\it\thesectionc.\thesubsectionc. #1}\par\vspace*{0.4cm}}
\renewcommand{\subsubsection}[1]
{\vspace*{0.6cm}\addtocounter{subsubsectionc}{1}
        \noindent
{\normalsize\rm\thesectionc.\thesubsectionc.\thesubsubsectionc.
        #1}\par\vspace*{0.4cm}}

\newcounter{appendixc}
\newcounter{subappendixc}[appendixc]
\newcounter{subsubappendixc}[subappendixc]

\renewcommand{\appendix}[1] {\vspace*{0.6cm}
        \refstepcounter{appendixc}
        \setcounter{figure}{0}
        \setcounter{table}{0}
        \setcounter{equation}{0}
        \renewcommand{\thefigure}{\Alph{appendixc}.\arabic{figure}}
        \renewcommand{\thetable}{\Alph{appendixc}.\arabic{table}}
        \renewcommand{\theappendixc}{\Alph{appendixc}}
        \renewcommand{\theequation}{\Alph{appendixc}.\arabic{equation}}
        \noindent{\bf Appendix \theappendixc #1}\par\vspace*{0.4cm}}

\def\abstracts#1{{

\centering{\begin{minipage}{12.2truecm}\baselineskip=12pt\noindent
        \centerline{Abstract}\vspace*{0.3cm}
        \parindent=0pt #1
        \end{minipage}}\par}}
\def\xxx#1{{

\centering{\begin{minipage}{12.2truecm}\baselineskip=12pt\noindent
        \parindent=0pt #1
        \end{minipage}}\par}}

\renewenvironment{thebibliography}[1]
        {\begin{list}{\arabic{enumi}.}
        {\usecounter{enumi}\setlength{\parsep}{0pt}
\setlength{\leftmargin 1.25cm}{\rightmargin 0pt}
         \setlength{\itemsep}{0pt} \settowidth
        {\labelwidth}{#1.}\sloppy}}{\end{list}}

\newcounter{itemlistc}
\newcounter{romanlistc}
\newcounter{alphlistc}
\newcounter{arabiclistc}

\newcommand{\fcaption}[1]{
        \refstepcounter{figure}
        \setbox\@tempboxa = \hbox{\footnotesize Fig.~\thefigure. #1}
        \ifdim \wd\@tempboxa  6in
           {\begin{center}
        \parbox{6in}{\footnotesize\baselineskip=12pt Fig.~\thefigure. #1}
            \end{center}}
        \else
             {\begin{center}
             {\footnotesize Fig.~\thefigure. #1}
              \end{center}}
        \fi}

\newcommand{\tcaption}[1]{
        \refstepcounter{table}
        \setbox\@tempboxa = \hbox{\footnotesize Table~\thetable. #1}
        \ifdim \wd\@tempboxa  6in
           {\begin{center}
        \parbox{6in}{\footnotesize\baselineskip=12pt Table~\thetable. #1}
            \end{center}}
        \else
             {\begin{center}
             {\footnotesize Table~\thetable. #1}
              \end{center}}
        \fi}

\def\@citex[#1]#2{\if@filesw\immediate\write\@auxout
        {\string\citation{#2}}\fi
\def\@citea{}\@cite{\@for\@citeb:=#2\do
        {\@citea\def\@citea{,}\@ifundefined
        {b@\@citeb}{{\bf ?}\@warning
        {Citation `\@citeb' on page \thepage \space undefined}}
        {\csname b@\@citeb\endcsname}}}{#1}}

\newif\if@cghi
\def\cite{\@cghitrue\@ifnextchar [{\@tempswatrue
        \@citex}{\@tempswafalse\@citex[]}}
\def\citelow{\@cghifalse\@ifnextchar [{\@tempswatrue
        \@citex}{\@tempswafalse\@citex[]}}
\def\@cite#1#2{{$\null^{#1}$\if@tempswa\typeout
        {IJCGA warning: optional citation argument
        ignored: `#2'} \fi}}

 1
 1
 1

\font\ninerm=cmr9

\def\jpa#1#2#3{ {\it Jour. of Phys. A} {\bf #1} (19#3) #2}
\def\mpla#1#2#3{ {\it Mod. Phys. Lett. A} {\bf #1} (19#3) #2}
\def\nc#1#2#3{ {\it Nuovo Cim.} {\bf #1} (19#3) #2}
\def\nupb#1#2#3{ {\it Nucl. Phys. B} {\bf #1} (19#3) #2}
\def\plb#1#2#3{ {\it Phys. Lett. B} {\bf #1} (19#3) #2}
\def\pr#1#2#3{ {\it Phys. Rev.} {\bf #1} (19#3) #2}
\def\prd#1#2#3{ {\it Phys. Rev. D} {\bf #1} (19#3) #2}
\def\zpc#1#2#3{ {\it Z. Phys. C} {\bf #1} (19#3) #2}
\begin{document}
\centerline{\large \bf In a hot, chirally symmetric phase,}
\baselineskip=22pt
\centerline{\large \bf $\pi^0$ doesn't go into $2 \gamma$,
but $\pi^0 \sigma$ does}
\baselineskip=16pt
\vspace*{0.6cm}
\centerline{\large \it Robert D. Pisarski}
\baselineskip=13pt
\centerline{ Dept. of Physics}
\centerline{Brookhaven National Laboratory}
\baselineskip=12pt
\centerline{Upton, NY 11973, USA}
\centerline{e-mail: pisarski@bnl.gov}
\centerline{BNL preprint BNL-RP-954}
\vspace*{0.3cm}
\vspace*{0.9cm}
\abstracts{
In a constituent quark model at nonzero temperature,
the amplitude for
$\pi^0 \rightarrow 2 \gamma$ vanishes in a chirally symmetric
phase, while that for
$\pi^0 \sigma \rightarrow 2 \gamma$ does not.
}

\vspace*{.5cm}

\xxx{
To appear in {\it From thermal field theory to neural networks:
a day to remember Tanguy Altherr}, editors P. Aurenche, P. Sorba,
and G. Veneziano, World Scientific Publishing.
}

\normalsize\baselineskip=15pt
\setcounter{footnote}{0}
\renewcommand{\thefootnote}{\alph{footnote}}

\section{In Memoriam}

This paper is dedicated to
Tanguy Altherr.  It is not the subject
on which I spoke at the workshop in his memory, which occured about eight
months before I wrote this.  In the interim, I think that I've
stumbled onto something interesting.
I include it here as a gift to Tanguy, as the sort of thing
he might have done had he lived.

\section{Anomalies}

If one constructs the axial current for the fundamental fermion fields
of a gauge theory, the divergence of the axial current typically has
anomalies: the divergence is not just that given by the
classical equations of motion, but has quantum contributions
at one loop order.\cite{ra}
The anomaly is not altered by contributions
to higher loop order.\cite{rb}  Only the ultraviolet region of the one
loop integrals contributes to the anomaly, so it is reasonable
that the effects of a medium --- such as a thermal bath ---
do not modify the anomaly.\cite{rc,rd}

The basic point of this note is that while the form of the anomaly
in terms of the {\it fundamental} fields is unrenormalized by
a thermal bath, how the anomaly manifests itself in terms of
{\it effective} fields {\it does} change with temperature.
For simplicity I limit myself to nonzero temperature, but the
same conclusions hold for a fermi sea.
I consider $\pi^0 \rightarrow 2 \gamma$ as a prototypical
process, but similar
conclusions hold for
$\omega \rightarrow \rho \pi$,
$\omega \rightarrow \pi \pi \pi$,
and $\pi \pi \rightarrow KKK$.
This note is the first announcement of work in progress.\cite{re}

I work in a constituent quark model, with two
flavors and three colors of quark fields $\psi$.  I neglect the coupling
of quarks to gluons, so the color only contributes a factor
of $N_c = 3$ to a quark loop.
I include the coupling of quarks to photons, $A^\mu$,
and to mesons, $\Phi$.  The meson field
$\Phi = \sigma t_0 + i \vec{\pi} \cdot \vec{t}$,
with $\sigma$ a $0^+$ meson, $\vec{\pi}$ the $0^-$ pions,
and the flavor matrices are $t_0 = {\bf 1}/2$,
$tr(t^a t^b) = \delta^{a b}/2$.

Left and right handed quark fields are constructed by
using the projectors $P_{\ell,r} = (1 \mp \gamma^5)/2$,
$\psi_{\ell,r} = P_{\ell,r} \psi$;
I work in euclidean spacetime with a positive definite metric,
and take $(\gamma^5)^2 = 1$.
A chirally symmetric lagrangian is
\begin{equation}
{\cal L} \; = \;
\overline{\psi}_\ell  \not \!\! D \; \psi_\ell
\; + \;
\overline{\psi}_r \not \!\! D \psi_r
\; + \; 2 \, \widetilde{g} \left(
\overline{\psi}_\ell \Phi \psi_r +
\overline{\psi}_r \Phi^\dagger \psi_\ell \right) \; .
\end{equation}
$\not \!\!\! D = (\not \!\! \partial
- i q  \not \!\!\! A )$, where $q$ is a matrix
for the electric charge of the up and down quarks,
$q = e (t_3 + t_0/3)$.
Excluding the electromagnetic
coupling, this lagrangian is
manifestly invariant under
global $SU_\ell(2) \times SU_r(2)$ chiral rotations $\Omega_\ell$
and $\Omega_r$,
\begin{equation}
\psi_{\ell,r} \rightarrow \Omega^\dagger_{\ell,r} \psi_{\ell,r}
\;\;\; , \;\;\;
\Phi \rightarrow \ \Omega_\ell^\dagger \,
\Phi \, \Omega_r \; .
\end{equation}
Explicitly,
\begin{equation}
{\cal L} \; = \;
\overline{\psi} \left( \not \!\! D \; + \; 2 \, \widetilde{g}
\left( \sigma t_0 \; + \; i \,
\vec{\pi} \cdot \vec{t} \gamma^5 \right)
\right) \psi \; .
\end{equation}

I do not consider the dynamics of either the scalar or quark
fields.  All I will do is
to derive the effective lagrangian between the
scalar and photon fields which is induced by
integrating out the quarks at one loop order.

I assume that chiral symmetry breaking occurs at zero
temperature, $\langle \sigma \rangle = \sigma_0$, so the
consituent quark mass is $m = \widetilde{g} \sigma_0$.
In a sigma model with two flavors, $\sigma_0 = f_\pi$, where
$f_\pi  = 93 \, MeV$ is the pion decay constant.

The Feynman rules required are the following: the
quark propagator is $1/(i \not \!\! P + m)$, the
coupling between $\sigma$ and a quark line is $- \widetilde{g}$,
the coupling between a pion $\pi^a$ and a quark line is
$- 2 i \widetilde{g} t^a \gamma^5$, and the coupling between
a photon and a quark line is $+ i q \gamma^\mu$.

\section{$\pi^0 \rightarrow 2 \gamma$ at zero temperature}

In this section I calculate
$\pi^0 \rightarrow 2 \gamma$ in a constituent quark model.\cite{ra,rf}
Let the two photons be $A^\mu(P_1)$ and $A^\nu(P_2)$, where
$P_1$ and $P_2$ are the four momenta.  There are two triangle
diagrams which contribute; one is
\begin{equation}
- \frac{i \widetilde{g} e^2 N_c}{3} \;
tr_K \; tr_{Dirac} \; \left( \; \gamma^5
\; \frac{1}{( i ( \not \!\! K + \not \!\! P_1 ) + m)} \; \gamma^\mu \;
\frac{1}{ ( i \not \!\! K + m)} \; \gamma^\nu \;
\; \frac{1}{(i (\not \!\! K - \not \!\! P_2) + m)} \right) \; .
\end{equation}
$tr_K = \int d^4 K/(2 \pi)^4$
is the integral over the loop momenta $K$, and
$tr_{Dirac}$ the Dirac trace.  The latter is done using the
identity
\begin{equation}
tr \left( \gamma^5 \gamma^\mu \gamma^\nu \gamma^\alpha
\gamma^\beta \right) \; = \; 4 \, \epsilon^{\mu \nu \alpha \beta}
\; ,
\end{equation}
with $\epsilon^{\mu \nu \alpha \beta}$ the antisymmetric tensor.
Doing the Dirac algebra, this diagram becomes
\begin{equation}
- \; \frac{4 i \widetilde{g} e^2 N_c}{3} \; m  \;
\epsilon^{\mu \nu \alpha \beta} P_1^\alpha P_2^\beta \;
I(P_1,P_2,m) \; ,
\end{equation}
where $I(P_1,P_2,m)$ is the loop integral
\begin{equation}
I(P_1,P_2,m) \; = \;
tr_K \;
\; \frac{1}{(K^2 + m^2) ( (K + P_1)^2 + m^2) ((K- P_2)^2 + m^2)}
\; .
\end{equation}
This integral is finite and well defined.  In the limit of small
momenta, the dependence on $P_1$ and $P_2$ in the integral can be
neglected, with
\begin{equation}
I(0,0,m) \; = \;
tr_K \; \frac{1}{(K^2 + m^2)^3} \; = \;
\frac{1}{32 \pi^2 m^2} \; .
\end{equation}

The second diagram, which follows by interchanging
$P_1$ and $P_2$, and $\mu$ and $\nu$, contributes equally.
Putting all of this together, the effective lagrangian for
$\pi^0 \rightarrow 2 \gamma$ is
\begin{equation}
{\cal L}_{\pi^0 \rightarrow 2 \gamma}
\; = \; + \, i \; \frac{e^2 N_c}{96 \pi^2 f_\pi}
\; \pi^0 \;
\epsilon^{\mu \nu \alpha \beta}
\; F^{\mu \nu} \; F^{\alpha \beta} \; .
\end{equation}
To derive this, I have used the relation
$m = \widetilde{g} f_\pi$.
To anticipate the results in the following sections, I note
that after integration by parts,
\begin{equation}
{\cal L}_{\pi^0 \rightarrow 2 \gamma}
\; = \; - \, i \; \frac{e^2 N_c}{24 \pi^2 f_\pi^2}
\; \left( f_\pi \partial^\mu \pi^0 \right) \;
\epsilon^{\mu \nu \alpha \beta}
\; A^\nu \; \partial^\alpha \; A^\beta \; .
\end{equation}

Given the derivation, it is clear that this is only the leading term
in an expansion in low momentum.  The integral
$I(P_1,P_2,m)$ can be
expanded in powers of the external momenta, in powers of
$P_1^2/m^2, P_2^2/m^2$, and so on.  In an effective lagrangian
these terms would becomes powers of $\partial^2/m^2$ acting
upon the various fields.
Thus at least for the couplings between pions and photons,
there is no such thing as
``the'' anomaly.  The lowest term,
${\cal L}_{\pi^0 \rightarrow 2 \gamma}$,
{\it is} unique in that it is
part of the Wess-Zumino-Witten term, whose form
can be derived from very general considerations,\cite{rg} apart from
an overall constant proportional to the number of
colors, $N_c$.
But even just from the integral at one loop order,
there will still be an infinity of couplings
between $\pi^0$ and
$\epsilon^{\mu \nu \alpha \beta}
F^{\mu \nu} F^{\alpha \beta}$; because of the
presence of the antisymmetric tensor, these couplings
can be uniquely identified with the anomaly.

\section{$\pi^0 \rightarrow 2 \gamma$ at non zero temperature}

Now consider the calculation of
$\pi^0 \rightarrow 2 \gamma$ in a thermal bath,
at a nonzero temperature $T$.  This is simply a matter of
computing the integral $I(P_1,P_2,m)$ at nonzero temperature.
I assume that the theory is in a phase near a point at
which chiral symmetry is restored, so that the
constituent quark mass $m$ is much less than the temperature.
Thus $m$ can be neglected in the integral, as can
$P_1$ and $P_2$.  The integral is that
for a fermion loop, so the timelike component of
the momentum $k^0 = (2n + 1) \pi T$, with a sum over all
integers $n$.  To do the integral it is easiest to integrate
over the spatial momentum $k$ first and then do the
sum over $n$:
$$
tr_K \; \frac{1}{(K^2)^3} \; = \;
T \; \sum_{n = - \infty}^{+ \infty} \; \int
\frac{d^3 k}{(2 \pi)^3} \;
\frac{1}{(k^2 + (k_0)^2)^3}
$$
\begin{equation}
\; = \; \frac{1}{16 \pi ^4 T^2} \;
\sum_{n = 1}^{\infty} \frac{1}{(2 n - 1)^3} \; = \;
\; \frac{7 \zeta(3)}{128 \pi^4 T^2} \; .
\end{equation}
$\zeta(3) = \sum^\infty_{n = 1} 1/n^3 = 1.20206...$ is
a type of zeta function, $\zeta(r) = \sum^\infty_{n = 1} 1/n^r$.

In the limit that $m \ll T$,
at nonzero temperature
the effective lagrangian for $\pi \rightarrow 2 \gamma$ is
then equal to
\begin{equation}
{\cal L}_{\pi^0 \rightarrow 2 \gamma}(T \neq 0)
\; = \; - i \;
\frac{7 \zeta(3) e^2 \widetilde{g}^2 N_c}{96 \pi^4 T^2}
\; \left(\sigma_0 \partial^\mu \pi^0 \right)
\; \epsilon^{\mu \nu \alpha \beta}
A^\nu \; \partial^\alpha \; A^\beta \; .
\end{equation}

This demonstrates that in a chirally symmetric phase, where
$\sigma_0 = 0$,
the amplitude for $\pi^0 \rightarrow 2 \gamma$
{\it vanishes}!
Actually, we could easily have anticipated that,
contrary to previous authors,\cite{rh}
the form of
the coupling between $\pi^0 \rightarrow 2 \gamma$ {\it must}
change at nonzero temperature.
At zero temperature, this
amplitude
is proportional to $1/f_\pi$.
Assume, for the purposes of argument,
that the restoration is a second order phase transition, at which
$f_\pi$ vanishes smoothly.  If the form of this coupling didn't
change, it would diverge, which is nonsensical for a physical
amplitude.

This doesn't explain why the
amplitude for $\pi \rightarrow 2 \gamma$ {\it vanishes}
in a chirally symmetric phase.  I
have written the above expression in a suggestive form.  The
coupling between the $\Phi$ field and photons must be gauge
invariant; as a term induced by the anomaly, it should also
involve the antisymmetric tensor.
With the total sigma field equal to $\sigma_0 + \sigma$,
the contribution of the
scalar fields to the axial current is
\begin{equation}
\vec{J}_{axial}^\mu \; = \; ( \sigma_0 + \sigma)
\partial^\mu \vec{\pi} - \vec{\pi} \partial^\mu \sigma \; .
\end{equation}
Thus the expression for
${\cal L}_{\pi^0 \rightarrow 2 \gamma}(T \neq 0)$ could simply
be the first term in the expression
\begin{equation}
{\cal L}_{\pi^0 \sigma \rightarrow 2 \gamma}(T \neq 0)
\; = \; - i \;
\frac{7 \zeta(3) e^2 \widetilde{g}^2 N_c }{96 \pi^4 T^2}
\; J^\mu_{3 \, , \, axial}
\; \epsilon^{\mu \nu \alpha \beta}
A^\nu \; \partial^\alpha \; A^\beta \; ,
\label{ef}
\end{equation}
$J^\mu_{3 \, , \, axial} = 2 \, tr(t^3
\vec{J}^\mu_{axial})$.

If correct, Eq.~(\ref{ef}) predicts that while the
amplitude for $\pi^0 \rightarrow 2 \gamma$ vanishes in a
chirally symmetric phase, that for $\pi^0 \sigma \rightarrow
2 \gamma$ does {\it not}.  It is relatively easy to check this.
In all, there are six diagrams which contribute.  One diagram is
\begin{equation}
\frac{i e^2 \widetilde{g}^2 N_c}{3} \;
tr_K \; tr_{Dirac} \; \left(
\frac{\gamma^5 \; \not \!\! K \;\gamma^\mu
\left( \not \!\! K - \not \!\! P_1 \right) \; \gamma^\nu \;
\left( \not \!\! K - \not \!\! P_1 - \not \!\! P_2\right) \;
\left( \not \!\! K + \not \!\! P_4 \right) }
{K^2 \left( K - P_1 \right)^2 \left( K + P_4 \right)^2
\left( K - P_1 - P_2 \right)^2 } \right) \; .
\end{equation}
As before,
the two photons are $A^\mu(P_1)$ and $A^\nu(P_2)$; the
momentum of the $\sigma$ is $P_3$, that of the pion $P_4$.
For arbitrary momenta the Dirac trace in this expression involves
trace of $\gamma^5$ times six $\gamma$'s, which is involved.
To simplify the Dirac algebra I assume that $P_4 = 0$.
(I checked that the result obtained for $P_3 = 0$ has the proper
change in sign as predicted from Eq.~(\ref{ef}).)  After doing
the Dirac algebra, the diagram becomes
\begin{equation}
 \frac{- 4 i e^2 \widetilde{g}^2 N_c}{3} \;
tr_K \; \left( \frac{\epsilon^{\mu \nu \alpha \beta} K^\alpha P_2^\beta}
{K^2 \left( K + P_1 \right)^2 \left( K - P_2 \right)^2} \right)
\; .
\end{equation}
The denominators can be expanded in small momenta, such as
\begin{equation}
\frac{1}{\left( K + P_1 \right)^2 }
\; = \; \frac{1}{K^2} \; - \;
\frac{2 K \cdot P_1}{\left( K^2 \right)^2} \; + \ldots \; .
\end{equation}
Remember that at nonzero temperature, the fermion loop momenta
$k^0$ is an odd multiple of $\pi T$, while
the bosonic momenta
$p^0$ are even multiples of $\pi T$.  In order to
be able to expand in powers of the external momenta, then,
implicitly it is necessary to assume that $p^0 = 0$ for each
of the external momenta.  Consequently, one of the indices
$\mu$ or $\nu$ must be time like.

The integral
\begin{equation}
tr_K \; \frac{K^\alpha K^\beta}{\left(K^2 \right)^4}
\; = \; \left( \delta^{\alpha \beta} + 2 n^\alpha n^\beta \right)
\; \frac{7 \zeta(3)}{128 \pi^4 T^2} \;
\end{equation}
is required, where $n^\alpha = \delta^{\alpha 0}$.  Actually, the
term proportional to $n^\alpha$ doesn't contribute, because
of the assumption that $n \cdot P_1 = n \cdot P_2 = 0$.
The diagram then equals
\begin{equation}
+ \; i \; \frac{7 \zeta(3) e^2 \widetilde{g}^2 N_c }
{288 \pi^4 T^2} \;
\epsilon^{\mu \nu \alpha \beta} \; P_1^\alpha \; P_2^\beta \; .
\end{equation}
This is only one of six diagrams.  All diagrams contribute
equally, so in all the amplitude is six times this expression.
This agrees with Eq.~(\ref{ef}), after remembering that because
there are two $A^\mu$'s, there
is a factor of two which arises in going from the effective
lagrangian to the amplitude.

A general understanding of how anomalous interactions induced
by fermion loops manifest themselves in a chirally symmetric phase
at nonzero temperature can be guessed from
the results for $\pi^0 \rightarrow 2 \gamma$.
I define an anomalous interaction as one which involves
the antisymmetric tensor, $\epsilon^{\mu \nu \alpha \beta}$.
This is then contracted with (gauge invariant) functions of
the external gauge fields and the scalar field, $\Phi$.
Rather obviously, in a chirally symmetric phase the
scalar field $\Phi$ must enter in a chirally invariant
manner.  In general, the operators so constructed have
a mass dimension greater than four.  At zero temperature, the proper
mass dimension is provided by adding powers of $1/f_\pi$;
in a chirally symmetric phase
at nonzero temperature, powers of $1/T$ enter.
Lastly, the expression must be invariant under
parity, under which $\Phi \leftrightarrow \Phi^\dagger$.

For example, the
axial current is the difference of left and right handed
currents,
\begin{equation}
J^\mu_{3 \, , \, axial} \; = \;
i \; tr \left( t^3 \left(
\Phi \partial^\mu \Phi^\dagger
\; - \; \partial^\mu \Phi \Phi^\dagger
+ \partial^\mu \Phi^\dagger \Phi
\; - \; \Phi^\dagger \partial^\mu \Phi \right) \right) \; .
\end{equation}
The reason why $\pi^0 \rightarrow 2 \gamma$ vanishes in a chirally
symmetric phase, then, is simply because there is no chirally
invariant function which is linear in the pion
field.  The simplest
chirally invariant function is bilinear in $\Phi$, so
the interaction begins as $\pi^0 \sigma \rightarrow 2 \gamma$.

Similar conclusions hold for other anomalous interactions.
All interactions of the $\omega$ meson are
anomalous:
at zero temperature the dominant couplings of the
$\omega$ are
\begin{equation}
\frac{1}{f_\pi} \;
\epsilon^{\mu \nu \alpha \beta} \; \omega^\mu \;
\partial^\nu
\vec{\rho}^{\, \alpha} \cdot \partial^\beta \vec{\pi} \; ,
\end{equation}
which governs $\omega \rightarrow \rho \pi$ and
\begin{equation}
\frac{1}{f_\pi^3} \;
\epsilon^{\mu \nu \alpha \beta} \; \omega^\mu \;
\partial^\nu \vec{\pi} \cdot
\left( \partial^\alpha \vec{\pi} \times
\partial^\beta \vec{\pi} \right) \; ,
\end{equation}
which describes $\omega \rightarrow \pi \pi \pi$.

Using the rules described above, in a chirally
symmetric phase at nonzero temperature the analogous
interactions are
\begin{equation}
\frac{1}{T^2} \; \epsilon^{\mu \nu \alpha \beta}
\; \omega^\mu \; \partial^\nu \vec{\rho}^{\, \alpha}
\cdot \; \vec{J}^{\, \beta}_{axial} \;
\end{equation}
and
\begin{equation}
\frac{1}{T^4} \; \epsilon^{\mu \nu \alpha \beta}
\; \omega^\mu \; tr \left(
\Phi^\dagger \; \partial^\nu \Phi \;
\partial^\alpha \Phi^\dagger \; \partial^\beta \Phi \; - \;
\Phi \; \partial^\nu \Phi^\dagger \;
\partial^\alpha \Phi \; \partial^\beta \Phi^\dagger \right) \;  .
\end{equation}
There are dimensionless coefficients in front of each expression,
which involve the various coupling constants and pure numbers,
such as $\zeta(3)$ and $\zeta(5)$, respectively.
The form of the second expression is dictated by parity invariance,
as the $\Phi$ dependent term must be odd under
$\Phi \leftrightarrow \Phi^\dagger$.  From
the form of these expressions, it is evident
that $\omega \rightarrow \rho \pi$ and
$\omega \rightarrow \pi \pi \pi$ vanish in a chirally symmetric
phase, but that the processes
$\omega \rightarrow \rho \pi \sigma$ and
$\omega \rightarrow \pi \pi \pi \sigma$ do not.
Under the standard assumption of vector meson dominance,\cite{rj}
this suggests that the width of the $\omega$
does not significantly increase with temperature.

The generalization of the Wess-Zumino-Witten term to nonzero temperature
can also be derived by these arguments.
At zero temperature, the Wess-Zumino-Witten term governs processes
such as $KK \rightarrow \pi \pi \pi$.
In a chirally symmetric phase at nonzero temperature,
$K K \rightarrow \pi \pi \pi$ vanishes, but
$K K \rightarrow \pi \pi \pi \sigma$ is allowed.
The extension of the Wess-Zumino-Witten term is
not a simple generalization of that at zero temperature,
as has been previously conjectured.\cite{ri}

\section{References}

\medskip
\end{document}